\documentstyle[preprint,aps]{revtex}

\newcommand{\be}{\begin{equation}}
\newcommand{\ee}{\end{equation}}
\newcommand{\Dlt}{\Delta}
\newcommand{\prt}{\partial}
\newcommand{\br}{{\bf r}}
\newcommand{\bk}{{\bf k}}

\newcommand{\al}{\alpha}
\newcommand{\ra}{\rightarrow}
\newcommand{\sgm}{\sigma}
\newcommand{\lbd}{\lambda}

\tightenlines

\begin{document}

\draft

\title{No Anomalous Fluctuations Exist in Stable Equilibrium Systems}

\author{V.I. Yukalov}

\address{Institut f\"ur Theoretische Physik, \\
Freie Universit\"at Berlin, Arnimallee 14, D-14195 Berlin, Germany \\
and \\
Bogolubov Laboratory of Theoretical Physics, \\
Joint Institute for Nuclear Research, Dubna 141980, Russia}

\maketitle

\begin{abstract}

A general theorem is rigorously proved for the case, when an observable 
is a sum of linearly independent terms: {\it The dispersion of a global 
observable is normal if and only if all partial dispersions of its terms 
are normal, and it is anomalous if and only if at least one of the partial 
dispersions is anomalous}. This theorem, in particular, rules out the 
possibility that in a stable system with Bose-Einstein condensate some 
fluctuations of either condensed or noncondensed particles could be 
anomalous. The conclusion is valid for arbitrary systems, whether uniform 
or nonuniform, interacting weakly or strongly. The origin of fictitious 
fluctuation anomalies, arising in some calculations, is elucidated.

\end{abstract}

\vskip 0.5cm


The problem of fluctuations of observable quantities is among the most 
important questions in statistical mechanics, being related to the 
fundamentals of the latter. A great revival, in recent years, of interest 
to this problem is caused by intensive experimental and theoretical studies 
of Bose-Einstein condensation in dilute atomic gases (see e.g. reviews
[1--3]). The fact that the ideal uniform Bose gas possesses anomalously
large number-of-particle fluctuations has been known long ago [4,5], which 
has not been of much surprise, since such an ideal gas is an unrealistic
and unstable system. However, as has been recently suggested in many papers, 
similar anomalous fluctuations could appear in real interacting Bose systems. 
A number of recent publications has addressed the problem of fluctuations in 
Bose gas, proclaiming controversial statements of either the existence or 
absence of anomalous fluctuations (see discussion in review [6]). So that 
the issue has not been finally resolved. In the present paper, the problem 
of fluctuations is considered from the general point of view, independent of 
particular models or calculational methods. A general theorem is rigorously 
proved, from which it follows that there are no anomalous fluctuations in any 
stable equilibrium systems.

It is worth stressing that no phase transitions are considered in this 
paper. As can be easily inferred from any textbook on thermodynamics or
statistical mechanics, the points of phase transitions are, by definition, 
the points of instability. A phase transition occurs exactly because one 
phase becomes unstable and has to change to another stable phase. It is 
well known that at the points of second-order phase transitions fluctuations 
do become anomalous, yielding divergent susceptibilities, as it should be at 
the points of instability. After a phase transition has occurred, the system, 
as is also well known, becomes stable and susceptibilities go finite. However,
in many papers on Bose systems, the claims are made that fluctuations remain 
anomalous far below the condensation point, in the whole region of the 
Bose-condensed system. As is shown below, these claims are incorrect, since 
such a system with anomalous fluctuations possesses a {\it divergent 
compressibility}, thus, being unstable.

Observable quantities are represented by Hermitian operators from the algebra 
of observables. Let $\hat A$ be an operator from this algebra. Fluctuations
of the related observable quantity are quantified by the dispersion
\be
\label{1}
\Dlt^2(\hat A) \equiv \;  <\hat A^2>\; - \; <\hat A>^2 \; ,
\ee
where $<\ldots>$ implies equilibrium statistical averaging. The dispersion 
itself can be treated as an observable quantity, which is the average of an 
operator $(\hat A-<\hat A>)^2$, since each dispersion is directly linked to 
a measurable quantity. For instance, the dispersion for the number-of-particle
operator $\hat N$ defines the isothermal compressibility
\be
\label{2}
\kappa_T \equiv -\; \frac{1}{V}\left ( \frac{\partial V}{\partial P}
\right )_T = \frac{\Dlt^2(\hat N)}{N\rho k_B T}  \; ,
\ee
in which $P$ is pressure, $\rho\equiv N/V$ is density, $N=<\hat N>$, $V$ 
is volume, and $T$ temperature. The dispersion $\Dlt^2(\hat N)$ is also 
connected with the sound velocity $s$ through the equation
\be
\label{3}
s^2 \equiv \frac{1}{m} \left ( \frac{\prt P}{\prt \rho} \right )_T =
\frac{1}{m\rho\kappa_T} = \frac{Nk_B T}{m\Dlt^2(\hat N)} \; ,
\ee
where $m$ is particle mass, $P$ pressure, and with the central value
\be
\label{4}
S(0) =\rho k_B T \kappa_T = \frac{k_BT}{ms^2} = \frac{\Dlt^2(\hat N)}{N}
\ee
of the structural factor 
$$
S(\bk)=1+\rho\int[g(\br)-1]e^{-i\bk\cdot\br}d\br \; ,
$$
where $g(\br)$ is the pair correlation function. The fluctuations of the
Hamiltonian $\hat H$ characterize the specific heat
\be
\label{5}
C_V \equiv \frac{1}{N} \left ( \frac{\prt E}{\prt T}\right )_V = 
\frac{\Dlt^2(\hat H)}{Nk_B T^2} \; ,
\ee
where $E\equiv<\hat H>$ is internal energy. In magnetic systems, with the
Zeeman interaction $-\mu_0\sum_i{\bf B}\cdot {\bf S}_i$, the longitudinal
susceptibility
\be
\label{6}
\chi_{\al\al} \equiv \frac{1}{N} \left ( \frac{\prt M_\al}{\prt B_\al}
\right ) = \frac{\Dlt^2(\hat M_\al)}{Nk_B T}
\ee
is related to the fluctuations of the magnetization $M_\al\equiv<\hat M_\al>$,
where $\hat M_\al\equiv\mu_0\sum_{i=1}^N S_i^\al$.

All relations (2) to (6) are exact and hold true for any equilibrium system. 
The stability conditions for such systems require that at any finite 
temperature, {\it except the points of phase transitions}, quantities (2) to 
(6) be positive and finite for all $N$, including the thermodynamic limit, 
when $N\ra\infty$. At the phase transition points, these quantities can of 
course be divergent, since, as is well known, the phase transition points 
are the points of instability. Summarizing this, we may write the general 
form of the {\it necessary stability condition} as
\be
\label{7}
0 < \frac{\Dlt^2(\hat A)}{N} < \infty \; ,
\ee
which must hold for any stable equilibrium systems at finite temperature and
for all $N$, including the limit $N\ra\infty$. The value $\Dlt^2(\hat A)/N$
can become zero only at $T=0$. Condition (7) is nothing but a representation 
of the well known fact that the susceptibilities in stable systems are
positive and finite.

The stability condition (7) shows that the dispersion $\Dlt^2(\hat A)$ has to 
be of order $N$. When $\Dlt^2(\hat A)\sim N$, one says that the dispersion is
normal and the fluctuations of an observable $\hat A$ are normal, since then 
the stability condition (7) is preserved. But when $\Dlt^2(\hat A)\sim N^\al$, 
with $\al>1$, then such a dispersion is called anomalous and the fluctuations 
of $\hat A$ are anomalous, since then $\Dlt^2(\hat A)/N\sim N^{\al-1}\ra\infty$
as $N\ra\infty$, hence the stability condition (7) becomes broken. A system 
with anomalous fluctuations is unstable. For example, an ideal uniform Bose 
gas, with $\Dlt^2(\hat N)\sim N^2$, is unstable [4--6].

Thus, in any stable equilibrium system, the fluctuations of global observables
must be normal, $\Dlt^2(\hat A)\sim N$. The situation becomes more involved,
if the operator of an observable is represented by a sum
\be
\label{8}
\hat A = \sum_i \hat A_i
\ee
of Hermitian terms $\hat A_i$, and one is interested in the fluctuations of
the latter. Then the intriguing question is: Could some partial dispersions
$\Dlt^2(\hat A_i)$ be anomalous, while the total dispersion $\Dlt^2(\hat A)$
remaining normal? Exactly such a situation concerns the systems with Bose 
condensate. Then the total number of particles $N=N_0+N_1$ is a sum of the  
numbers of condensed, $N_0$, and noncondensed, $N_1$, particles. We know that 
for a stable system $\Dlt^2(\hat N)$ must be normal. But could it happen that
at the same time either $\Dlt^2(\hat N_0)$ or $\Dlt^2(\hat N_1)$, or both
would be anomalous, as is claimed by many authors?

Considering the sum (8), it is meaningful to keep in mind a nontrivial case,
when all $\hat A_i$ are linearly independent. In the opposite case of linearly
dependent terms, one could simply express one of them through the others and
reduce the number of terms in sum (8). The consideration also trivilizes if 
some of $\hat A_i$ are $c$-numbers, since then $\Dlt^2(c)=0$.

The dispersion of operator (8) reads as
\be
\label{9}
\Dlt^2(\hat A) = \sum_i \Dlt^2(\hat A_i) + 2 \sum_{i<j} {\rm cov}
(\hat A_i,\hat A_j) \; ,
\ee
where the covariance
$$
{\rm cov}(\hat A_i,\hat A_j) \equiv \frac{1}{2}\; <\hat A_i\hat A_j +
\hat A_j\hat A_i>\; - \; <\hat A_i><\hat A_j> 
$$
is introduced. The latter is symmetric, ${\rm cov}(\hat A_i,\hat A_j)=
{\rm cov}(\hat A_j,\hat A_i)$. The dispersions are, by definition, non-negative,
but the covariances can be positive as well as negative. One might think that
an anomalous partial dispersion $\Dlt^2(\hat A_i)$ could be compensated by some
covariances, so that the total dispersion $\Dlt^2(\hat A)$ would remain normal.
This is just the way of thinking when one finds an anomalous dispersion of 
condensed, $\Dlt^2(\hat N_0)$, or noncondensed, $\Dlt^2(\hat N_1)$, particles,
presuming that the system as a whole could remain stable, with the normal total
dispersion $\Dlt^2(\hat N)$. However, the following theorem rules out such
hopes.

\vskip 3mm

{\bf Theorem}. The total dispersion (9) of an operator (8), composed of 
linearly independent Hermitian operators, is anomalous if and only if at
least one of the partial dispersions is anomalous, with the power of the
total dispersion defined by that of its largest partial dispersion.
Conversely, the total dispersion is normal if and only if all partial 
dispersions $\Dlt^2(\hat A_i)$ are normal.

\vskip 2mm

{\bf Proof}. First of all, we notice that it is sufficient to prove the
theorem for the sum of two operators, for which
\be
\label{10}
\Dlt^2(\hat A_i + \hat A_j)= \Dlt^2(\hat A_i) + \Dlt^2(\hat A_j) +
2\; {\rm cov}(\hat A_i,\hat A_j) \; ,
\ee
where $i\neq j$. This is because any sum of terms more than two can always be
represented as a sum of two new terms. Also, we assume that both operators in
Eq. (10) are really operators but not $c$-numbers, since if at least one of
them, say $\hat A_j=c$, is a $c$-number, then Eq. (10) reduces to a simple
equality $\Dlt^2(\hat A_i+c)=\Dlt^2(\hat A_i)$ of positive (or semipositive)
quantities, both of which simultaneously are either normal or anomalous.

Introduce the notation
\be
\label{11}
\sgm_{ij} \equiv {\rm cov}(\hat A_i,\hat A_j) \; .
\ee
As is evident, $\sgm_{ii}=\Dlt^2(\hat A_i)\geq 0$ and $\sgm_{ij}=\sgm_{ji}$.
The set of elements $\sgm_{ij}$ forms the covariance matrix $[\sgm_{ij}]$,
which is a symmetric matrix. For a set of arbitrary real-valued numbers,
$x_i$, with $i=1,2,\ldots,n$, where $n$ is an integer, one has
\be
\label{12}
< \left [ \sum_{i=1}^n \left (\hat A_i -\; <\hat A_i> \right )
x_i \right ]^2 > \; = \sum_{ij}^n \sgm_{ij} x_i x_j \; \geq 0 \; .
\ee
The right-hand side of equality (12) is a semipositive quadratic form. From 
the theory of quadratic forms [7] one knows that a quadratic form is 
semipositive if and only if all principal minors of its coefficient matrix 
are non-negative. Thus, the sequential principal minors of the covariance 
matrix $[\sgm_{ij}]$, with $i,j=1,2,\ldots, n$, are all non-negative. In 
particular, $\sgm_{ii}\sgm_{jj}-\sgm_{ij}\sgm_{ji}\geq 0$. This, owing to 
the symmetry $\sgm_{ij}=\sgm_{ji}$, transforms to the inequality 
$\sgm_{ij}^2\leq\sgm_{ii}\sgm_{jj}$. Then the correlation coefficient
\be
\label{13}
\lbd_{ij} \equiv \sgm_{ij}/\sqrt{\sgm_{ii}\sgm_{jj}}
\ee
possesses the property $\lbd_{ij}^2\leq 1$.

The equality $\lbd_{ij}^2=1$ holds true if and only if $\hat A_i$ and $\hat 
A_j$ are linearly dependent. The sufficient condition is straightforward, 
since if $\hat A_j=a+b\hat A_i$, where $a$ and $b$ are any real numbers, then
$\sgm_{ij}=b\sgm_{ii}$ and $\sgm_{jj}=b^2\sgm_{ii}$, hence $\lbd_{ij}=b/|b|$,
from where $\lbd_{ij}^2=1$. To prove the necessary condition, assume that
$\lbd^2_{ij}=1$. This implies that $\lbd_{ij}=\pm 1$. Consider the dispersion
$$
\Dlt^2\left ( \frac{\hat A_i}{\sqrt{\sgm_{ii}}} \pm 
\frac{\hat A_j}{\sqrt{\sgm_{jj}}} \right ) = 2 (1\pm \lbd_{ij}) \geq 0 \; .
$$
The value $\lbd_{ij}=1$ is possible then and only then, when 
$$
\Dlt^2 \left ( \frac{\hat A_i}{\sqrt{\sgm_{ii}}}\; - \;
\frac{\hat A_j}{\sqrt{\sgm_{jj}}} \right ) = 0 \; .
$$ 
The dispersion can be zero if and only if 
$$
\frac{\hat A_i}{\sqrt{\sgm_{ii}}}\; -\; 
\frac{\hat A_j}{\sqrt{\sgm_{jj}}} = const \; ,
$$ 
that is, the operators $\hat A_i$ and $\hat A_j$ are linearly dependent. 
Similarly, the value $\lbd_{ij}=-1$ is possible if and only if 
$$
\frac{\hat A_i}{\sqrt{\sgm_{ii}}} +
\frac{\hat A_j}{\sqrt{\sgm_{jj}}} = const \; ,
$$ 
which again means the linear dependence of the operators $\hat A_i$ and 
$\hat A_j$. As far as these operators are assumed to be linearly independent, 
one has $\lbd^2_{ij}<1$. The latter inequality is equivalent to 
$\sgm_{ij}^2<\sgm_{ii}\sgm_{jj}$, which, in agreement with notation (11), 
gives
\be
\label{14}
|{\rm cov}(\hat A_i,\hat A_j)|^2 < \Dlt^2(\hat A_i)\Dlt^2(\hat A_j) \; .
\ee

The main relation (10) can be written as
\be
\label{15}
\Dlt^2(\hat A_i +\hat A_j) = \sgm_{ii} + \sgm_{jj} +
2\lbd_{ij}\sqrt{\sgm_{ii}\sgm_{jj}} \; ,
\ee
where, as is shown above, $|\lbd_{ij}|<1$. Altogether there can occur no more
than four following cases. First, when both partial dispersions $\sgm_{ii}=
\Dlt^2(\hat A_i)$ and $\sgm_{jj}=\Dlt^2(\hat A_j)$ are normal, so that 
$\sgm_{ii}\sim N$ and $\sgm_{jj}\sim N$. Then from Eq. (15) it is evident 
that the total dispersion $\Dlt^2(\hat A_i + \hat A_j)\sim N$ is also normal. 
Second, one of the partial dispersions, say $\sgm_{ii}\sim N$, is normal, but 
another is anomalous, $\sgm_{jj}\sim N^\al$, with $\al>1$. From Eq. (15), 
because of $(1+\al)/2<\al$, one has $\Dlt^2(\hat A_i+\hat A_j)\sim N^\al$, 
so that the total dispersion is anomalous, having the same power $\al$ as
$\sgm_{jj}$. Third, both partial dispersions are anomalous, $\sgm_{ii}\sim 
N^{\al_1}$ and $\sgm_{jj}\sim N^{\al_2}$, with different powers, say $1<\al_1 
<\al_2$. From Eq. (15), taking into account that $(\al_1+\al_2)/2<\al_2$, we 
get $\Dlt^2(\hat A_i+\hat A_j)\sim N^{\al_2}$, that is, the total dispersion  
is also anomalous, with the same power $\al_2$ as the largest partial 
dispersion $\sgm_{jj}$. Fourth, both partial dispersions are anomalous, 
$\sgm_{ii}=c_i^2N^\al$ and $\sgm_{jj}=c_j^2N^\al$, where $c_i>0$ and 
$c_j>0$, with the same power $\al$. Then Eq. (15) yields 
$$
\Dlt^2(\hat A_i+\hat A_j)= c_{ij} N^\al
$$ 
with 
$$
c_{ij}=(c_i-c_j)^2+2c_ic_j(1+\lbd_{ij})>0 \; ,
$$
which is strictly positive in view of the inequality $|\lbd_{ij}|<1$. Hence, 
the total dispersion is anomalous, with the same power $\al$ as both partial 
dispersions. After listing all admissible cases, we see that the total 
dispersion $\Dlt^2(\hat A_i+\hat A_j)$ is anomalous if and only if at least 
one of the partial dispersions is anomalous, with the power of $N$ of the 
total dispersion being equal to the largest power of partial dispersions. 
Oppositely, the total dispersion is normal if and only if all partial 
dispersions are normal. This concludes the proof of the theorem.

\vskip 3mm

As an example, let us consider a Bose system with Bose-Einstein condensate,
whose total number-of-particle operator $\hat N=\hat N_0+\hat N_1$ consists
of two terms, corresponding to condensed, $\hat N_0$, and noncondensed,
$\hat N_1$, particles. Since for a stable system the dispersion $\Dlt^2(\hat 
N)$ is normal, then from the above theorem it follows that both dispersions 
$\Dlt^2(\hat N_0)$ as well as $\Dlt^2(\hat N_1)$ must be normal. No anomalous 
fluctuations can exist in a stable system, neither for condensed nor for 
noncondensed particles. This concerns any type of stable systems, either 
uniform or nonuniform. And this result does not depend on the method of 
calculations, provided the latter are correct.

How then could one explain the appearance of numerous papers claiming the 
existence of anomalous fluctuations in Bose-condensed systems in the whole 
region far below the critical point? If these anomalous fluctuations would 
really exist, then the compressibility would be divergent everywhere below 
the critical point. A system, whose compressibility is divergent everywhere
in its region of existence, as is known from any textbook on statistical
mechanics, is unstable. Such anomalous fluctuations are usually obtained 
as follows. One considers a low-temperature dilute Bose gas, at $T\ll T_c$, 
when the Bogolubov theory [8] is applicable. In the frame of this theory, 
one calculates the dispersion $\Dlt^2(\hat N_1)$ of noncondensed particles, 
where $\hat N_1=\sum_{k\neq 0}a_k^\dagger a_k$. To find $<\hat N_1^2>$, one 
needs to work out the four-operator expression $<a_k^\dagger a_k a_q^\dagger 
a_q>$, or after employing the Bogolubov canonical transformation $a_k=
u_kb_k+v_{-k}b_{-k}^\dagger$, to consider $<b_k^\dagger b_k b_q^\dagger b_q>$.
Such four-operator expressions are treated by invoking the Wick decoupling. 
Then one finds that the dispersion $\Dlt^2(\hat N_1)$ diverges as $N\int 
dk/k^2$. Discretizing the phonon spectrum, one gets $\Dlt^2(\hat N_1)\sim 
N^{4/3}$. Another way [6] could be by limiting the integration by the 
minimal $k_{min}=1/L$, where $L\sim N^{1/3}$ is the system length. Then 
again $\Dlt^2(\hat N_1)\sim N^{4/3}$. In any case, one obtains anomalous 
fluctuations of noncondensed particles. This result holds true for both 
canonical and grand canonical ensembles, which is a direct consequence of 
the Bogolubov theory [8]. But the anomalous behaviour of $\Dlt^2(\hat N_1)$, 
according to the theorem proved above, immediately leads to the same anomalous
behaviour of the total dispersion $\Dlt^2(\hat N)$, which would imply the 
instability of the system, since then compressibility (2) and structural
factor (4) become divergent. As far as the system is assumed to be stable, 
there should be something wrong in such calculations. 

The drawback of these calculations is in the following. One of the basic
points of the Bogolubov theory is in omitting in the Hamiltonian all terms
of orders higher than two with respect to the operators $a_k$ of noncondensed
particles. This is a {\it second-order} theory with respect to $a_k$. It means
that the same procedure of keeping only the terms of second order, but 
ignoring all higher-order terms, must be done in calculating any physical 
quantities. In considering $<\hat N_1^2>$, one meets the fourth-order terms 
with respect to $a_k$. Such fourth-order terms are not defined in the 
Bogolubov theory. The calculation of the fourth-order products in the
second-order theory is not self-consistent. This inconsistency leads to 
incorrect results.

A correct calculations of $\Dlt^2(\hat N)$ in the frame  of the Bogolubov 
theory can be done in the following way [9]. Writing down the pair 
correlation function, one should retain there only the terms not higher 
than of the second order with respect to $a_k$, omitting all higher-order 
terms. Then for a uniform system one gets
$$
g(\br) = 1 +\frac{2}{\rho} \int \left ( <a_k^\dagger a_k> + <a_k a_{-k}>
\right ) e^{i\bk\cdot\br}\; \frac{d\bk}{(2\pi^3)} \; .
$$
This gives us the structural factor $S(0)=k_B T/mc^2$, $c\equiv\sqrt{(\rho/m)
\Phi_0}$, $\Phi_0 =\int\Phi(\br)d\br$, where $\Phi(\br)$ is an interaction 
potential. Because of the exact relation $\Dlt^2(\hat N)=NS(0)$, we find the 
dispersion $\Dlt^2(\hat N)= N k_BT/mc^2$, which is, of course, normal, as it 
should be for a stable system. And since the total dispersion $\Dlt^2(\hat N)$
is normal, the theorem tells us that both the partial dispersions, $\Dlt^2(
\hat N_0)$ as well as $\Dlt^2(\hat N_1)$, must also be normal. Note that
if in defining the pair correlation function $g(\br)$ one would retain the 
higher-order terms, one would again get the anomalous total dispersion
$\Dlt^2(\hat N)$.

It is easy to show that the same type of fictitious anomalous fluctuations
appear for arbitrary systems, if one uses the second-order approximation 
for the Hamiltonian but intends to calculate fourth-order expressions. This 
is immediately evident from the analysis of susceptibilities for arbitrary 
systems with continuous symmetry, as is done by Patashinsky and Pokrovsky [10]
(Chapter IV), when the system Hamiltonian is restricted to hydrodynamic 
approximation. Following Ref. [10], one may consider an operator $\hat A=\hat 
A(\varphi)$ being a functional of a field $\varphi$. Let this operator be 
represented as a sum $\hat A=\hat A_0+\hat A_1$, in which the first term is 
quadratic in the field $\varphi$, so that $\hat A_0\sim\varphi^+\varphi$, 
while the second term depends on the field fluctuations $\delta\varphi$ as 
$\hat A_1\sim\delta\varphi^+\delta\varphi$. Keeping in the Hamiltonian only 
the second-order field fluctuations is equivalent to the hydrodynamic
approximation. The dispersion $\Dlt^2(\hat A)\sim N\chi$ is proportional to a 
{\it longitudinal} susceptibility $\chi$. The latter is given by the integral 
$\int C(\br)d\br$ over the correlation function $C(\br)=g(\br)-1$, where 
$g(\br)$ is a pair correlation function. Calculating $\Dlt^2(\hat A)$, one 
meets the fourth-order term $<\delta\varphi^+\delta\varphi\delta\varphi^+
\delta\varphi>$. If this is treated by invoking the Wick decoupling and the 
quadratic hydrodynamic Hamiltonian, one gets $C(\br)\sim 1/r^{2(d-2)}$ for any
dimensionality $d>2$. Consequently, $\chi\sim\int C(\br)d\br\sim N^{(d-2)/3}$ 
for $2<d<4$, and  the dispersion $\Dlt^2(\hat A)\sim N\chi\sim N^{(d+1)/3}$. 
For $d=3$, this results in the anomalous dispersion $\Dlt^2(\hat A)\sim N^{4/3}$. 
If this would be correct, this would mean, according to the necessary 
stability condition (7), that the system is unstable. That is, there could not
exist any stable systems with continuous symmetry, such as magnetic systems or
liquid helium. Of course, we know that such systems perfectly exist, but the 
above contradiction has arisen solely due to an inconsistent calculational 
procedure, when the fourth-order term $<\delta\varphi^+\delta\varphi\delta
\varphi^+\delta\varphi>$ was treated in the frame of the hydrodynamic 
approximation, which is a second-order theory with respect to $\delta\varphi$.

\vskip 3mm

Concluding, there are no anomalous fluctuations of any physical quantities 
in arbitrary stable equilibrium systems. Fictitious anomalous fluctuations
in realistic systems far outside any points of phase transitions  might 
appear only due to drawbacks in a calculational procedure. The absence of 
anomalous fluctuations follows from the general theorem, rigorously proved 
in this paper.

It is important to stress that not only the anomalous fluctuations as 
such signify the occurrence of instability, though they are the explicit 
signals of the latter. But also one should not forget that the dispersions
for the operators of observables are directly related to the corresponding
susceptibilities, as in Eqs. (2) to (6). It is the wrong behaviour of
these susceptibilities, which manifests the instability.

Thus, the anomalous number-of-particle fluctuations are characterized by 
the anomalous dispersion $\Dlt^2(\hat N)$. The latter is connected, through 
the general and exact relation (2), with the isothermal compressibility 
$\kappa_T$. The anomalous dispersion $\Dlt^2(\hat N)$ implies the divergence 
of this compressibility. From the definition of the compressibility 
$\kappa_T\equiv-(1/V)(\prt V/\prt P)$, it is obvious that its divergence 
means the following: An infinitesimally small positive fluctuation of
pressure abruptly squeezes the system volume to a point. Respectively, an
infinitesimally small negative fluctuation of pressure suddenly expands 
the system volume to infinity. It is more than evident that a system which 
is unstable with respect to infinitesimally small fluctuations of pressure,
immediately collapsing or blowing up, has to be termed unstable.

\vskip 5mm

{\bf Acknowledgement}

\vskip 2mm

I am very thankful to E.P. Yukalova for many discussions of mathematical 
and other problems. I am grateful to the German Research Foundation for the 
Mercator Professorship.

\end{document}